\documentclass[%
 reprint,
 amsmath,amssymb,
prb,
]{revtex4-2}
\usepackage{graphicx}% Include figure files
\usepackage{dcolumn}% Align table columns on decimal point
\usepackage{bm}% bold math
\usepackage{hyperref}
\usepackage[T1]{fontenc}
\usepackage[polish,english]{babel}
\usepackage[utf8]{inputenc}
\hypersetup{
    citecolor=blue,
    colorlinks=true,
    linkcolor=blue,
    filecolor=magenta,      
    urlcolor=blue,
}% add hypertext capabilities
\usepackage[colorinlistoftodos]{todonotes}
\usepackage[version=4]{mhchem}
\usepackage{multirow}

\begin{document}
\renewcommand{\figurename}{FIG.}
\renewcommand{\tablename}{TABLE}
\preprint{APS/123-QED}

\title{Me-graphane: tailoring the structural and electronic properties of Me-graphene by hydrogenation}% Force line breaks with \\

\author{Enesio Marinho Jr}
\email{enesio.junior@ufabc.edu.br}
%\author{Cedric Rocha Leão\,\href{https://orcid.org/0000-0002-1365-6514}{\includegraphics[scale=.05]{orcid_id}}}
\author{Pedro A. S. Autreto}
 \email{pedro.autreto@ufabc.edu.br}
\affiliation{%
Centro de Ciências Naturais e Humanas, Universidade Federal do ABC (UFABC),\\Avenida dos Estados 5001, 09210-580 Santo André, São Paulo, Brazil
}%


%

\date{December 11, 2020}% It is always \today, today,

\begin{abstract}
Graphene-based materials (GBMs) constitute a large family of materials which has attracted great interest for potential applications. In this work, we apply first-principles calculations based on density functional theory (DFT) and fully atomistic reactive molecular dynamics (MD) simulations to study the structural and electronic effects of hydrogenation in Me-graphene, a non-zero bandgap GBM composed of both $sp^2$ and $sp^3$-hybridized carbon. Our DFT results show a substantial tuning of the electronic properties of Me-graphene by hydrogenation, with the bandgap varying from $0.64$ eV to $2.81$ eV in the GGA-PBE approach, passing through metallic ground-states and a narrower bandgap state depending on the hydrogen coverage. %The effective electron mass is also modulated, ranging from 0.23 $m_0$ to 1.02 $m_0$ for fully-hydrogenated Me-graphene, but achieving the value of 0.14 $m_0$ for $15\% $-hydrogenated Me-graphene. 
The analyses of structural properties and binding energies have shown that hydrogenated Me-graphene presents strong and stable \ce{C-H} bonds, and all of the carbon atoms are in $sp^3$ hybridization resulting in a boat-like favorable conformation for fully-hydrogenated Me-graphene. Our MD simulations have indicated that the hydrogenation of Me-graphene is temperature-dependent, and the covalent adsorption tends to grow by islands. Those simulations also show that the most favorable site, predicted by our DFT calculations, acts as trigger adsorption for the extensive hydrogenation.
\end{abstract}

\maketitle

%\tableofcontents

\section{\label{sec:intro}Introduction}

%Graphene is indeed one of the most remarkable achievements in the condensed matter field. This two-dimensional (2D) allotrope of carbon is a one-atom-thick layer of $sp^2$ bonded carbon atoms arranged on a honeycomb structure made out of hexagons \cite{RMPCastroNeto}. Although planar graphene itself has been presumed not to exist in the free state, Novoselov \textit{et al.} successfully obtained it in 2004 using scotch tape to peel flakes of graphite off \cite{Novoselov2004}.  

Graphene \cite{Novoselov2004}, a two-dimensional (2D) allotrope of carbon arranged on a honeycomb structure made out of hexagons \cite{RMPCastroNeto}, has shown exceptional physical, chemical and mechanical properties. It has a large theoretical specific surface area (2,630 m$^2$\,g$^{-1}$) \cite{zhu2010graphene}, high electron mobility at room temperature (250,000 cm$^2$\,V$^{-1}$\,s$^{-1}$) \cite{novoselov2005two}, thermal conductivity in the order of 5000 W\,mK$^{-1}$ \cite{Balandin2008}, high Young’s modulus ($\sim$1.1 to 2.0 TPa) \cite{Lee385,JiangPRB2009}, and good electrical conductivity \cite{PAPAGEORGIOU201775,gluchowski2020}. More recently, Cao \textit{et al.} \cite{cao2018unconventional} experimentally demonstrated that bilayer graphene, which normally consists of two vertically stacked monolayer graphene layers arranged in an AB (Bernal) stacking configuration, when rotated at the so-called ``magic angle'' of $\sim$1.1° presents an intrinsic unconventional superconductivity for a critical temperature around 1.7 K. 

Owing to this impressive gamma of astonishing properties, graphene has demonstrated outstanding performance in several applications such as catalysis \cite{machado2012catalysis}, gas sensors \cite{yuan2013gassensors}, anti-corrosive coatings \cite{cui2019comprehensive}, flexible touchscreens \cite{vlasov2017touchscreen}, supercapacitors \cite{wang2009supercapacitor}, transistors \cite{schwierz2010graphene}, spintronic \cite{han2014spintronics}, twistronics \cite{CarrPRBtwistronics}, energy storage \cite{olabi135energyStorage}, and most.  

However, the synthesis of high-quality and large-area monolayer graphene in a cost effective process represents a drawback for industrial-scale applications \cite{lin2019NM}. In addition, the unmodified graphene has certain limitations, such as weak electrochemical activity, easy agglomeration and difficult processing, which greatly limit its applications \cite{yu2020review}. Furthermore, this 2D nanomaterial is intrinsically a zero-gap semiconductor, or a semimetal,  which turn it unsuitable for switching devices \cite{pulizzi2019graphene}. This are some of the reasons for the huge increase in the number of researches focused on functionalization of graphene including reactions with organic and inorganic molecules, chemical modification of the large graphene surface, and the general description of various covalent and non covalent interactions with it \cite{georgakilas2012ChemRev}. Chemical functionalization of graphene  is a suitable approach to induce a tunable band gap opening \cite{pumera2014heteroatom}, or to enable this material to be processed by solvent-assisted techniques, such as layer-by-layer assembly, spin-coating, and filtration \cite{kuila2012chemical}.

The chemical functionalization by attaching hydrogen adatoms forming a fully hydrogenated graphene changes the hybridization of carbon atoms from $sp^2$ to $sp^3$, thus removing the conducting $\pi$-bands and opening a direct band gap at the $\Gamma$ point with magnitude of $3.5$-$3.7$ eV \cite{sofo2007PRBgraphane}. The so-called graphane is nonmagnetic semiconductor composed of 100\% hydrogenated graphene, resulting in a CH stoichiometry. This graphene-based material was predicted to be stable in an extended covalently bonded 2D hydrocarbon with two favorable conformations: chair-like conformer, with the hydrogen atoms alternating on both sides of the plane and a boat-like conformer with the hydrogen adatoms alternating in pairs \cite{sofo2007PRBgraphane}. Graphane was first synthesized by Elias \textit{et al.} using free-standing graphene, and further the authors have shown that reaction with hydrogen is reversible, so that the original metallic state, the lattice spacing, and even the quantum Hall effect can be restored by annealing \cite{elias2009Science}.   

Another important stable graphene derivative was already obtained using fluorine adatoms that strongly bind to carbons given rise to the so-called fluorographene, a 2D fully fluorinated graphene analogue of a fully fluorinated one-dimensional carbon chain known as Teflon.  Fluorographene is a wide band gap semiconductor with $E_{\text{g}}\,{=}\,3.8$ eV,  wide enough for optoelectronic applications in the blue/UV spectrum \cite{nair2010fluorographene,jeon2011fluorographene}. 
\begin{figure*}[t]
    \centering
    \includegraphics[width=\textwidth]{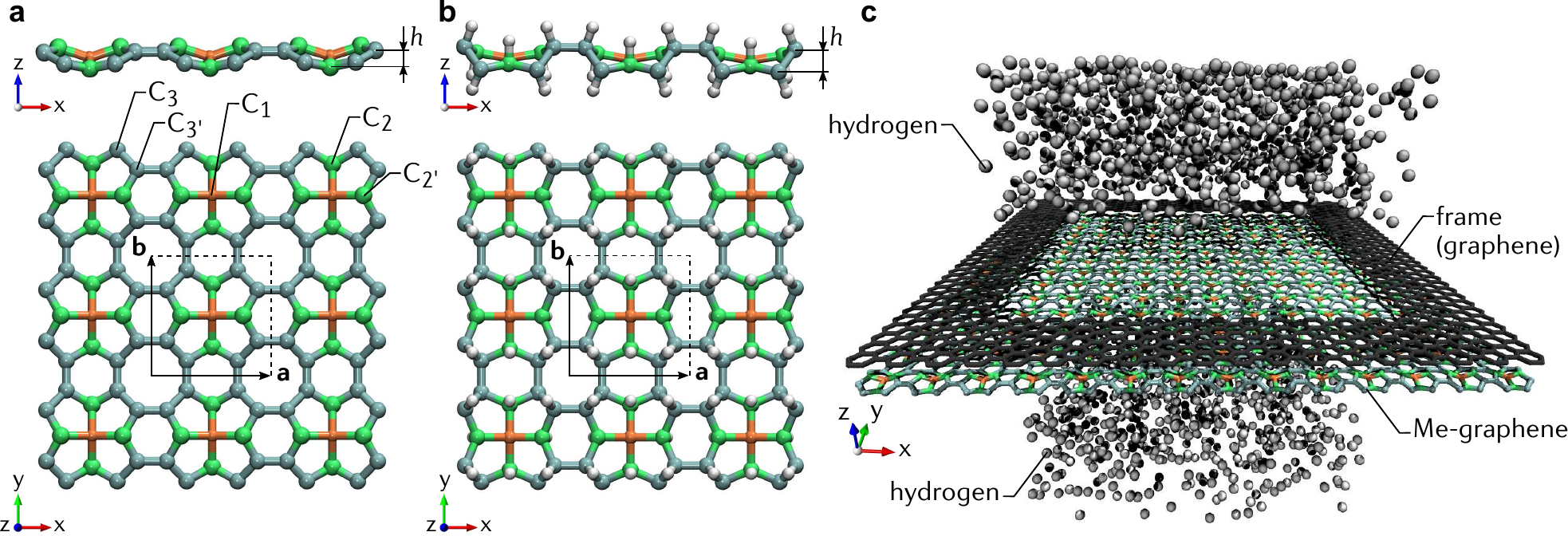}
    \caption{Top and side view of optimized structure of (a) Me-graphene and (b) Me-graphane. We describe three types of carbon atoms, labeled as C$_1$ ($sp^3$) (orange), C$_2$/C$_{2'}$ ($sp^2$) (green) and C$_3$/C$_{3'}$ ($sp^2$) (blue). For Me-graphane, the hydrogen atoms are represented by white spheres. The buckling height is denoted by $h$. The unit cells are represented by doted square and lattice vectors $\mathbf{a}$ and $\mathbf{b}$. (c) Representative structure used in the molecular dynamics simulations of the hydrogenation process of Me-graphene, in which we included a frame of graphene to prevent bending or folding of the membrane.}
    \label{fig:me-graph_structure}
\end{figure*}
Janus graphene (J-GN) has been predicted theoretically and prepared experimentally to study asymmetric chemistry of graphene functionalization \cite{yang2013JGN,li2015JGN,zhang2013JGN}. The J-GN is prepared achieving asymmetric covalent functionalization with a variety of functional groups on the opposite sides of graphene. The hydrofluorinated J-GN, namely fluorographone, is a semiconducting graphene derivative formed by covalent functionalization with hydrogen and fluorine adatoms being adsorbed onto the opposite sides of monolayer graphene \cite{jin2016SciRepJGN}. Recently, we studied the modulation of hydrogen adsorption and the corresponding variation in electronic properties of hydrofluorinated J-GN \cite{PRM-Schleder-Marinho2020}.

The innovation arising out of graphene has boosted and inspired the research and development of novel 2D materials graphene-like allotropes such as graphyne's family \cite{baughman1987structure},   biphenylene carbon (BPC, also called graphenylene) \cite{baughman1987structure, brunetto2012nonzero, enyashin2011graphene}, penta-graphene \cite{PuruJena2015P-GN}, pentahexoctite \cite{sharma2014pentahexoctite}, T-graphene \cite{liu2012structural}, octagraphene \cite{sheng2012octagraphene}. % among others \cite{bhimanapati2015recent,alvarez2017adsorption,paupitz2012graphene}.
The carbon atoms in these allotropes are either $sp^2$ and/or $sp$ hybridized. Interestingly, all these allotropes are planar and exhibit unique electronic as well
as mechanical properties \cite{sharma2014pentahexoctite}. 

Zhuo \textit{et al.} \cite{zhuo2020M-GN} have predicted a new dynamically stable 2D carbon allotrope named tertiary-methyl-graphene or Me-graphene (M-GN). This 2D carbon-based material is composed of both $sp^2$ and $sp^3$-hybridized carbon by topological assembly of \ce{C-(C3H2)4} molecules. Consisting of a transitional ratio of $sp^2$ and $sp^3$-hybridized carbon atoms at the ratio of $12:1$, M-GN is a transition between graphene (ratio $1:0$) and penta-graphene (ratio $2:1$). As expected, M-GN has transition properties between those of graphene and penta-graphene. For example, its band gap of 1.08 eV, which is between graphene (semimetal) and penta-graphene ($E_{\text{g}}\,{=}\,3.25$ eV \cite{PuruJena2015P-GN}). Furthermore, M-GN presents an unusual near zero Poisson’s ratio of $-0.002$ up to $0.009$ in the xy-plane, different from that of graphene ($0.169$) and penta-graphene ($-0.068$). M-GN also exhibits a high hole mobility of $1.60{\times} 10^5$ cm$^2$\,V$^{-1}$\,s$^{-1}$ at 300 K \cite{zhuo2020M-GN}. Those interesting properties of M-GN can be tuned by chemical functionalization, adsorbing for example hydrogen adatoms onto its surfaces. Indeed, this route tends to be explored in the future in order to open up new possibilities of applications for functionalized graphene-based materials.   

Herein, we carried out a comprehensive atomistic study on the structural and electronic properties of M-GN covalent functionalized with hydrogen. For this purpose, we performed \textit{ab initio} calculations based on density functional theory, as well as fully atomistic reactive molecular dynamics within the ReaxFF force field. We predicted drastic changes in electronic and structural properties of M-GN by hydrogenation, pushing forward the frontiers of novel semiconducting 2D materials based on covalent functionalization of graphene. 

\section{\label{sec:method} COMPUTATIONAL DETAILS}
\begin{figure*}[t]
    \centering
    \includegraphics[width=\textwidth]{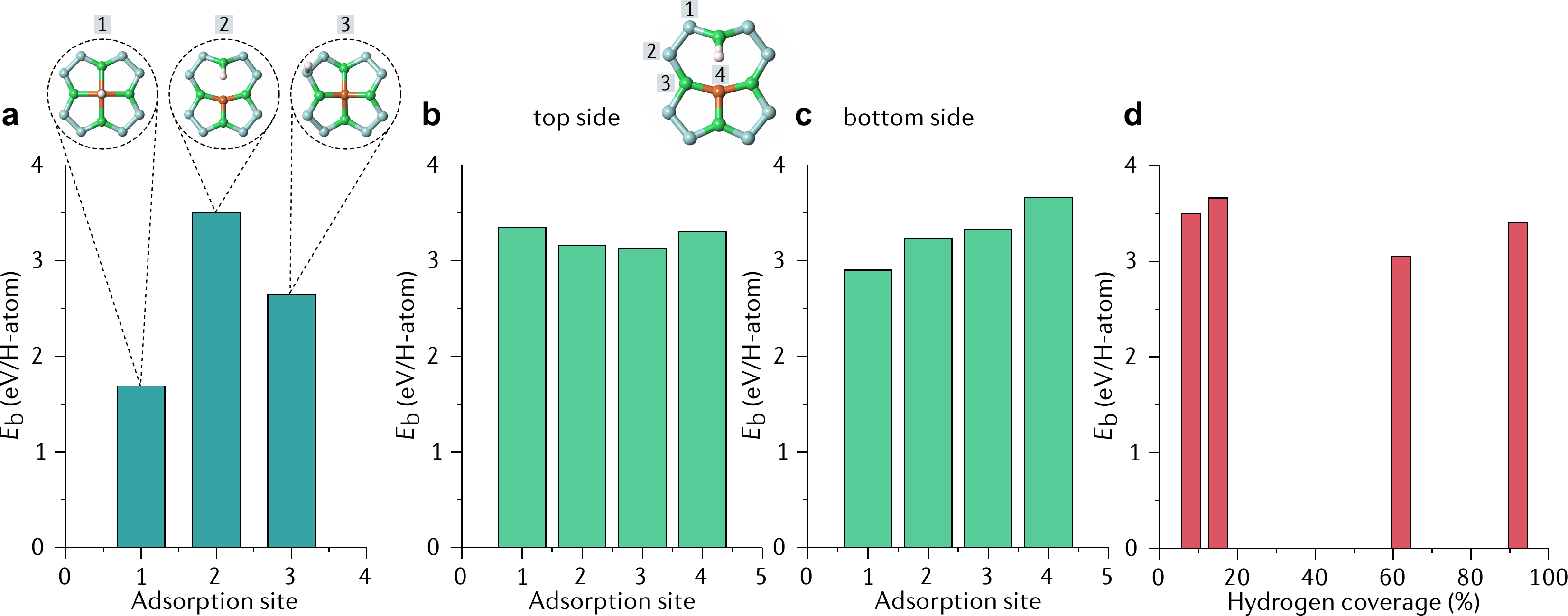}
    \caption{Binding energies for hydrogenation of Me-graphene: (a) single-hydrogenated, (b) double-hydrogenated on top side and (c) bottom side, and (d) binding energies per hydrogen adatom as function of hydrogen coverage. The upper panels show the respective adsorption sites in Me-graphene.}
    \label{fig:binding-energies}
\end{figure*}
We performed \textit{ab initio} calculations based on density functional theory \cite{HohenbergKohn1964,KohnSham1965} as implemented in the Vienna ab initio simulation package (\textsc{vasp}) \cite{VASP}. The projector augmented-wave method \cite{PAW} was used to treat the electron-ion interaction, and the exchange-correlation functional was described by the generalized gradient approximation (GGA) as proposed by Perdew, Burke, and Ernzerhof (PBE) \cite{PBE}. Structural optimizations were performed using a conjugate gradient algorithm until the residual forces on atoms reach values smaller than $0.025$ eV\,\AA$^{-1}$. Kohn-Sham orbitals were expanded into a plane-wave basis set with kinetic energy up to 500 eV. The Brillouin zone was sampled using a $\Gamma$-centered $17{\times}17{\times}1$ $k$-point mesh for the structural optimization of M-GN, following the scheme proposed by Monkhorst and Pack \cite{Monkhorst-Pack}. M-GN monolayer and its images were separated by a vacuum space of $\sim20$ \AA\, to avoid spurious interactions.

The crystal structure of M-GN is formed by twelve $sp^2$-hybridized and one $sp^3$-hybridized carbon atoms, as illustrated in Fig.~\ref{fig:me-graph_structure}\textcolor{blue}{(a)}, which can be described as methane with four hydrogen atoms replaced by cyclocopropenylidenes \cite{zhuo2020M-GN}. The ground-state structure posses a symmetry of P$\bar{4}m$2 (space group \#115), containing carbon pentagons, hexagons, and octagons. The representative structure of fully-hydrogenated M-GN, named Me-graphane, is shown in Fig.~\ref{fig:me-graph_structure}\textcolor{blue}{(b)}. The modification of the $sp^2$-$sp^2$ carbon bonds by formation of $sp^3$-derived carbon-hydrogen bonds lead to the variation of structural conformation.

 In addition, we carried out fully atomistic molecular dynamics (MD) simulations to study hydrogen adsorption and structural properties of hydrogenated M-GN at different temperatures. We employed the reactive force field (ReaxFF), which is an empirical force field for reactive systems developed by van Duin \textit{et al.} \cite{vanDuinReaxFF}. ReaxFF employs a bond order/bond energy relationship, which allows for bond formation and bond dissociation during molecular dynamics (MD) simulations. The bond orders are obtained from interatomic distances and are updated at every MD or energy minimization step \cite{raju2013reaxff}. ReaxFF MD simulations were implemented using the large-scale atomic/molecular massively parallel simulator (\textsc{lammps}) code \cite{lammps}. Precisely, we applied the well-established \ce{C-C} interaction parameters developed by Chenoweth \textit{et al.} \cite{chenoweth2008reaxff}. All ReaxFF MD simulations have been performed in the canonical (NVT) ensemble, with a time step of $0.25$ fs using the Nosé-Hoover thermostat with a coupling time constant of $25$ fs to control the temperature of the entire system. The hydrogenation was carried out in a suspended M-GN membrane supported by a graphene frame as shown in Fig.~\ref{fig:me-graph_structure}\textcolor{blue}{(c)}. The hydrogen atmosphere was composed of $1{,}000$ atoms in a volume of $58{,}300$ \AA$^3$ on both sides of the membrane. We also restricted the hydrogenation in the suspended Me-G area to avoid edge effects. 
 
\section{\label{sec:results} Results and Discussions}

\subsection{Structural properties of the pristine and hydrogenated Me-graphene}
\begin{figure*}[t]
    \centering
    \includegraphics[width=.9\textwidth]{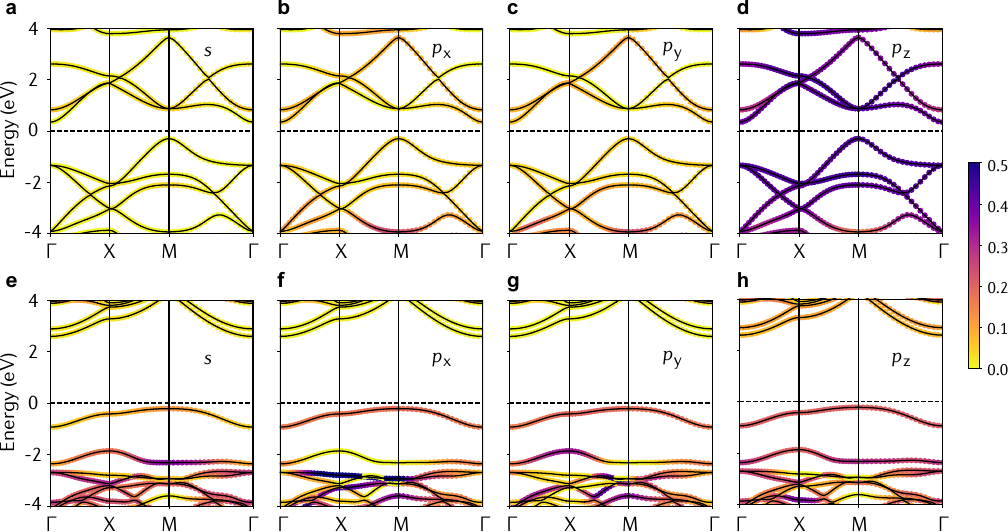}
    \caption{Orbital-resolved bandstructures for the atomic orbitals (a) $s$, (b) $p_x$, (c) $p_y$ and (d) $p_z$ of Me-graphene, and for the orbitals (e) $s$, (f) $p_x$, (g) $p_y$ and (h) $p_z$ of Me-graphane. The scale indicates the magnitude of the projection. The Fermi level was set to zero of energy.}
    \label{fig:bands_orbitals}
\end{figure*}
We analyze the structural modifications of M-GN due to hydrogen adsorption onto its surface. In Table~\ref{tab:struc-parameters} we present the optimized lattice constants of M-GN, regarding $a$ and $b$ lattice vectors and buckling height $h$. We obtained $a=b=5.745$ \AA, with a thickness of $h=0.985$ \AA, in close agreement with Ref.~\cite{zhuo2020M-GN} ($a=b=5.744$ \AA\, and $h=0.988$ \AA). We have estimated three types of bond length between the $sp^2$-hybridized carbons, with values of $1.416$, $1.449$ and $1.406$ \AA, similar to the \ce{C-C} bond length in graphene (1.426 Å). Between $sp^3$-hybridized (\ce{C1}) and $sp^2$-hybridized (\ce{C2}) carbons, the bond length was estimated in 1.563 \AA. The bond angle among \ce{C1-C2-C3} is 114.2 degrees, larger than the 109.47 degrees for standard $sp^3$ hybridization angle, and that one formed among \ce{C2-C1-C_{2'}} is 95.71 degrees. 
\begin{table}[t]
    \centering
    \caption{Structural parameters for Me-graphene and Me-graphane.}
    {\def\arraystretch{1.2}\tabcolsep=3.5pt
    \begin{tabular}{lccc}
    \hline\hline
         & & Me-graphene & Me-graphane \\\hline
         \multirow{3}{*}{\shortstack[l]{Lattice \\parameters (\AA)}} & $a$ & 5.745 & 5.904 \\
                                           & $b$ &5.745 &5.905 \\
         & $h$ & 0.985 &1.110 \\[3pt]
        \multirow{3}{*}{\shortstack[l]{Bond \\lengths (\AA)}} &\ce{C1-C2} & 1.563 &1.614\\ 
                          &\ce{C2-C3} & 1.416 &1.530\\
                          &\multirow{2}{*}{\ce{C3-C$_{3'}$}} &1.449 &1.579\\
                          & &1.406 &1.549\\[3pt]
                          
        \multirow{2}{*}{\shortstack[l]{Bond \\angles ($^{\circ}$)}} &\ce{C1-C2-C3} &114.22 &114.16\\ 
                          &\ce{C2-C1-C$_{2'}$} &95.71 &91.59\\\hline\hline
    \end{tabular}
    }
    \label{tab:struc-parameters}
\end{table}

Comparing the results of M-GN to those obtained for Me-graphane, we notice that the hydrogenation promotes an overall increase of the structural parameters, inducing a lattice strain of $\sim2.8\%$. Every \ce{C-C} bond length in Me-graphane is higher or similar to $sp^3$ bond length of $1.548$ \AA\, in diamond, and much greater than $1.426$ \AA\, characteristic of $sp^2$ \ce{C-C} bond in graphene \cite{sofo2007PRBgraphane,zhuo2020M-GN}. We find that the boat-like conformer of Me-graphane is the most favorable conformation, with a buckling height of $1.110$ \AA. Similarly, the \ce{C-C} bond lengths in graphane are $1.52$ \AA, much similar to that in diamond and shorter than \ce{C-C} bond lengths in graphene. Furthermore, graphane  has two favorable conformations, which are chair-like and boat-like conformers \cite{sofo2007PRBgraphane}.

\subsection{Hydrogen binding energy}

We evaluate the most stable adsorption sites
for hydrogenation calculating the hydrogen binding energies for different sites as the difference in total energy between the hydrogenated-compounds and their component parts as follows:
\begin{equation}
    E_{\text{b}} = -\left[\frac{E_{\text{M-GN$+n$H}}-\left(E_{\text{M-GN}}+E_{\text{$+n$H}}\right)}{n}\right]\,,
\end{equation}
where $E_{\text{M-GN$+n$H}}$ is the ground-state energy for the M-GN containing a total of $n$ adsorbed hydrogen atoms,  as well as $E_{\text{M-GN}}$ and $E_{\text{$n$H}}$ are the energies for pristine M-GN and for an isolated hydrogen times the total number of hydrogen adsorbed, respectively. The results of binding energy per hydrogen in M-GN are shown in Fig.~\ref{fig:binding-energies}. The adsorption of one H into \ce{C2} site (index 2 in figure) of M-GN, corresponding to $7\%$ of hydrogen coverage, leads to a 3.50 eV/H-atom binding energy (Fig.~\ref{fig:binding-energies}\textcolor{blue}{(a)}). When H is adsorbed in \ce{C2} site, the \ce{C1-C2} bond is broken, resulting in a decrease of the free energy due to a gain of conformational entropy yielding the highest binding energy. 

We also verify the hydrogen binding energy for a second adatom being inserted in single-hydrogenated M-GN, $8\%$ of hydrogen coverage (Fig.~\ref{fig:binding-energies}\textcolor{blue}{(b)-(c)}). Considering the H adsorption into the top side of M-GN, the highest binding energies were verified for the nearest-neighbor \ce{C3} (index 1 in figure) and for \ce{C1} site (index 4 in figure), although the H adsorption on the other two sites also results in high binding energies above 3 eV/H-atom. For the H adsorption in the bottom side, the highest binding energy was achieved for the \ce{C1} site. Finally, we study the binding energy for $62\%$-hydrogenated M-GN and for Me-graphane and, as a result, all the hydrogen binding energies were higher than 3 eV/H-atom. For comparison, the hydrogen binding energy in hydrogenated penta-graphene is 3.65 eV/H-atom \cite{li2016PCCP-hfpg}, slightly higher than the ones obtained for hydrogenated M-GN. Therefore, the hydrogenation of M-GN results in highly stable chemisorption and the resultant hydrogenated M-GN tends to be more thermodynamically favorable owing to the saturation of all nonplanar $sp^2$ hybridized carbon atoms.

\subsection{Electronic structures}

\begin{figure}[t]
    \centering
    \includegraphics[width=0.46\textwidth]{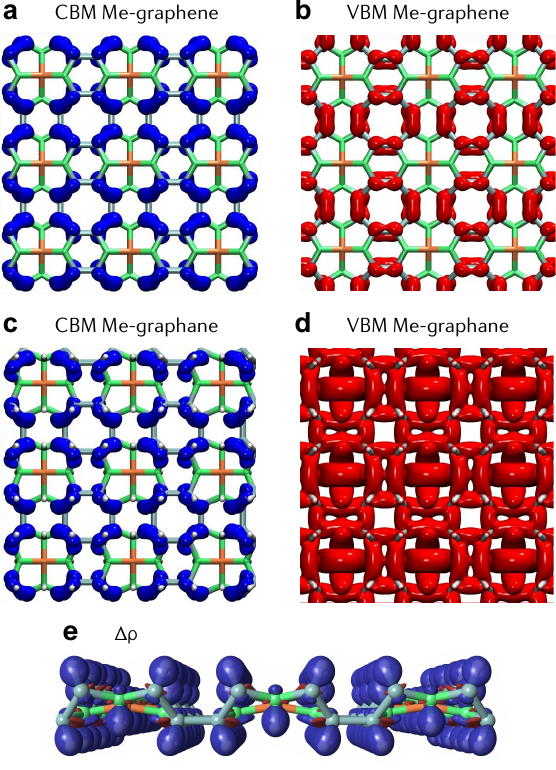}
    \caption{Partial charge densities of VBM and CBM for (a-b) Me-graphene and (c-d) Me-graphane, with isosurfaces value 0.005 $e$\,Å$^{-3}$. (e) Charge density difference ($\Delta \rho$) plot for Me-graphane with isosurface value 0.015 $e$\,Å$^{-3}$. In $\Delta \rho$ results, blue and red regions depict charge accumulation and depletion, respectively.}
    \label{fig:vbm-cbm-deltarho}
\end{figure}
In Fig.~\ref{fig:bands_orbitals}, we present the orbital-resolved band structures for M-GN and Me-graphane. The color scale indicates the magnitude of state has 4f or 5d character. Our results show that M-GN is an indirect band gap semiconductor with a band gap of 0.64 eV using GGA-PBE functional, with VBM at M ($0.5, 0.5, 0$) and CBM at $\Gamma$ point, in excellent agreement with the reported theoretical band gap of $0.65$ eV for M-GN \cite{zhuo2020M-GN} with identical band edges' coordinates in Brillouin zone. In M-GN, the states at both VBM and CBM have predominantly C-$p_z$ character, with no significant contributions from  C-$p_x$ and C-$p_y$ orbitas. On the other hand, the electronic band structure of Me-graphane indicate that this is a wide band gap 2D semiconductor, presenting an indirect band gap with magnitude of 2.81 eV in GGA-PBE approach with band edges located also at M (VBM) and $\Gamma$ (CBM). Near VBM, we notice the formation of a fully-filled intermediate band with bandwidth of about $1$ eV. The states of CBM in Me-graphane have mostly C-$p_z$ character analogous to M-GN, whereas VBM is mainly formed by hybridization of C-$p_x$, C-$p_y$ and C-$p_z$ orbitals. As a result, the intermediate band near VBM of Me-graphane occurs due to structural conformation of hydrogenated M-GN, with main contributions of carbon-related orbitals and not directly by the presence of adsorbed hydrogen atoms. 
\begin{table}[t]
    \centering
    \caption{Band gap and effective electron ($m^*_e$) and hole masses ($m^*_h$) in units of free electron mass $m_0$ at $\Gamma$ (CBM) and M (VBM) points, respectively, using GGA-PBE approach. Each effective mass was obtained from two high-symmetry directions in Brillouin zone: $\Gamma-$X and $\Gamma-$X for $m^*_e$, M$-$X and M$-\Gamma$ for $m^*_h$. }
    \label{tab:eg-and-masses}
    {\def\arraystretch{1.2}\tabcolsep=3.5pt
    \begin{tabular}{lcccccc}
    \hline\hline
         \multirow{2}{*}{H coverage (\%)} & \multirow{2}{*}{$E_{\text{g}}$ (eV)} & \multicolumn{2}{c}{$m^*_e/m_0$} & & \multicolumn{2}{c}{$m^*_h/m_0$} \\\cline{3-4}\cline{6-7}
         & & $\Gamma{-}$X& $\Gamma{-}$M & & M${-}$X & M${-}\Gamma$ \\\hline
         0 (Me-graphene) & 0.64 & 0.23 & 0.26 & & $-0.21$ & $-0.24$\\
         8 & 0 & -- & -- & & -- & --\\
         15 & 0.11 & 0.14 & 0.18 & &$-0.27$ & $-0.31$\\
         62 & 0 & -- & -- & & -- & --\\
         92 (Me-graphane)& 2.81 & 1.02 & 1.02 & & $-3.71$ &$-3.37$\\\hline\hline
    \end{tabular}
    }
\end{table}

The partial charge density distributions for CBM and VBM of M-GN and Me-graphane are shown in Fig.~\ref{fig:vbm-cbm-deltarho}\textcolor{blue}{(a)-(d)}. For M-GN, occupied-electron states of VBM and also empty CBM states are localized at octagonal rings. With regard to Me-graphane, CBM states lie in \ce{C3-C3} bonds of octagonal rings, whereas VBM states in Me-graphane are delocalized over all carbon bonds, in striking agreement with the orbital-resolved band structures and representing the hybridization of the C-$p_x$, C-$p_y$ and C-$p_z$ orbitals.
%%%
%$Delta \rho$
%%%
\begin{figure*}[t]
    \centering
    \includegraphics[width=\textwidth]{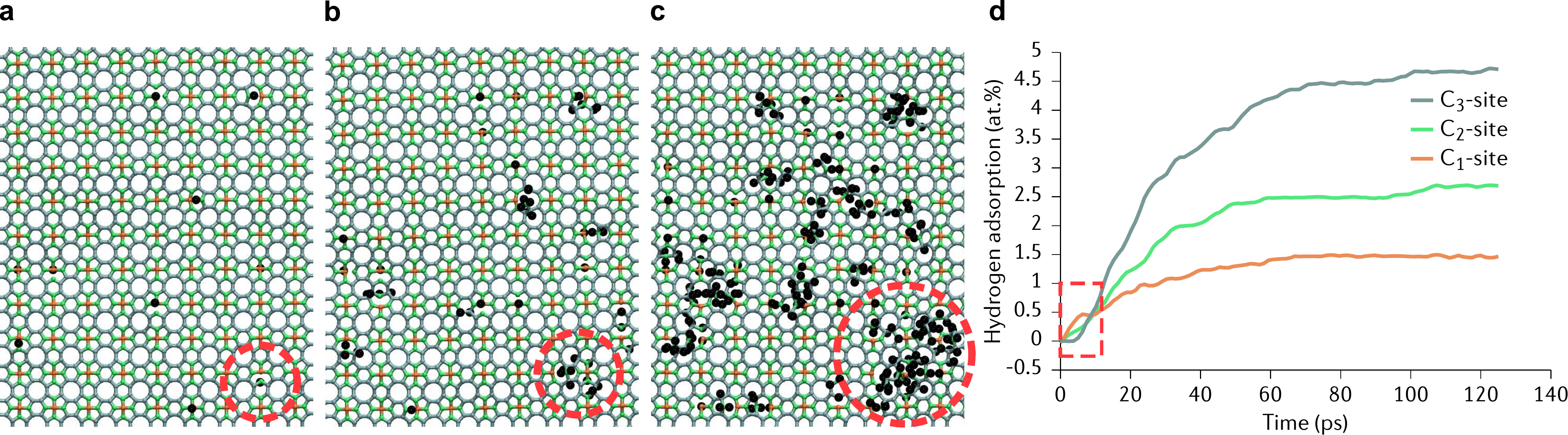}
    \caption{Representative snapshots from the reactive molecular dynamics simulations of hydrogen incorporation process, where black spheres represent H atoms. (a) Early stage, (b) intermediate stage, and (c) final stage of the H adsorption dynamics. (d) Hydrogen adsorption rate in atomic percent as a function of reaction time at 300 K, regarding the adsorption sites \ce{C1}, \ce{C2}, and \ce{C3}. }
    \label{fig:snaps-adscurve}
\end{figure*}

To describe the effects of hydrogenation in the electron distribution of functionalized M-GN relative to the unperturbed system, we compute the charge density difference as follows
\begin{equation}
    \Delta\rho = \rho_{\text{M-GN$+n$H}} - \left(\rho_{\text{M-GN}}+ \rho_{\text{$n$H}}\right)\,,
\end{equation}
where $\rho_{\text{M-GN$+n$H}}$ is the electron density of the hydrogenated M-GN containing a total of $n$ hydrogen adatoms, and $\rho_{\text{M-GN}}$ and $\rho_{\text{$n$H}}$ are the unperturbed electron densities of the substrate and sorbate, respectively. The $\Delta \rho$ result is shown in Fig.~\ref{fig:vbm-cbm-deltarho}\textcolor{blue}{(e)}. Our results indicate that $p$ orbitals of carbon lost a small amount of electron density, verified through the depletion volumes in red. We also notive that the hydrogen adatoms effectively gained electrons, as we can see in the accumulation volumes in blue around hydrogen atoms, matching with the expectation to form \ce{C-H} chemical bonds after adsorption.    

Furthermore, to analyze the tuning of electronic properties of M-GN by hydrogenation, we calculate the electronic band gap and effective electron and hole masses as a function of the amount of the hydrogen coverage, and the results are described in Table~\ref{tab:eg-and-masses}. The effective masses is a convenient approach to obtain quantitative insights on the mobility of charge carriers. The effective electron ($m^*_e$) and holes ($m^*_h$) masses were derived from parabolic fits to  the  GGA-PBE band  structures at the band extrema CBM located at $\Gamma$ and VBM located at M point, respectively, along the principal directions M${-}$X and M${-}\Gamma$ for VBM, and $\Gamma{-}$X and $\Gamma{-}$M for CBM. 

Our results indicate a dramatic variation of the band gap of M-GN by hidrogenation, ranging from 0.64 eV for pristine M-GN to 2.81 eV for Me-graphane. Moreover, for some intermediate concentrations of adsorbed hydrogen such as 8\% and 62\%, the hydrogenation produces metallic ground-state of functionalized M-GN with band gap being vanished. In turn, for $15\%$ of hydrogen coverage in M-GN we found a indirect band gap of 0.11 eV, which is 530 meV smaller than the band gap of pristine M-GN. This last hydrogenated M-GN system was modeled adsorbing one H on \ce{C2}-site (top side) and the other H onto nearest-neighbor \ce{C3}-site. 

The effective electron and hole masses were also effectively tailored with changes in hydrogen coverage of M-GN. The lowest values for pristine M-GN are similar, with $m^*_e = 0.23\,m_0$ in $\Gamma{-}$X direction and $m^*_h = 0.21\,m_0$ in M${-}$X. Comparing to penta-graphene which presents $m^*_e = 0.24\,m_0$ and $m^*_h = 0.50\,m_0$ using GGA-PBE functional \cite{deb2020P-GN}, the effective hole mass in M-GN tends to be lower while the effective electron mass is analogous. Our results for Me-graphane show that the effective masses significantly increase by full hydrogenation of M-GN. We compute $m^*_e = 1.02\,m_0$ and $m^*_h = 3.71\,m_0$ in GGA-PBE for Me-graphane. Similarly, the effective masses for fully-hydrogenated penta-graphene, named penta-graphane, also tends to be higher than those of pristine penta-graphene, with $m^*_e = 1.2\,m_0$ and $m^*_h = 0.58\,m_0$ applying GGA-revised-PBE functional \cite{einollahzadeh2016P-GN-H}. Conversely, the effective electron mass in 15\%-hydrogenated M-GN decreases to $m^*_e = 0.14\,m_0$ in $\Gamma{-}$X, with effective hole mass kept in the same order of magnitude, with $m^*_h = 0.27\,m_0$ in M${-}$X direction.            

\subsection{Reactive molecular dynamics}
The hydrogen adsorption dynamics was analyzed by reactive MD simulations. In Fig.~\ref{fig:snaps-adscurve}\textcolor{blue}{(a)-(c)} we present representative snapshots from
a $0.12$ ns MD simulation of the hydrogenation process in M-GN at 300 K for atmospheres composed only of H atoms inserted in both top and bottom side of the membrane. In the initial stage, Fig.~\ref{fig:snaps-adscurve}\textcolor{blue}{(a)}, the H atoms are mostly incorporated on \ce{C1} and \ce{C2}-sites. Verifying the snapshot for intermediate stage, Fig.~\ref{fig:snaps-adscurve}\textcolor{blue}{(b)}, hydrogen-adsorbed \ce{C2}-sites acts as seeds to the growth of hydrogen islands due to geometric changes caused by modification of carbon hybridization from $sp^2$ to $sp^3$. In Fig.~\ref{fig:snaps-adscurve}\textcolor{blue}{(d)} we show the curves of hydrogenation per total adsorption sites (at.\%) as a function of reaction time. At the beginning of hydrogenation, the H adsorption occurs mostly in \ce{C1} and \ce{C2}-sites of M-GN, which agrees with the snapshots (see dashed square and circles in Fig.~\ref{fig:snaps-adscurve}). 

In our reactive MD simulations, we have considered the hydrogen adsorption at temperatures of $150$, $300$, and $800$ K. As a result, the hydrogen rate incorporation can be represented in two phases, starting with a linear upward curve followed by a plateau indicating the saturation of adsorption and dynamic equilibrium. The saturation is achieved for different reaction times for each analyzed temperatures, suggesting therefore that the hydrogenation of M-GN is temperature-dependent reaction, which agrees with reported results of graphene's covalent functionalization \cite{paupitz2012graphene, PRM-Schleder-Marinho2020}. Our reactive MD simulations also show that the \ce{C1-C2} bond tends to be broken, favoring the formation of defects and turning the \ce{C1} site into a more favorable adsorption site.

\section{\label{sec:conclusion} Conclusions}

Motivated by the promising properties of a new graphene-based semiconductor named Me-graphene, we have studied the effects of hydrogenation on structural and electronic properties of this 2D nanomaterial. Our \textit{ab initio} DFT calculations show a extreme modulation of the electronic properties of M-GN by hydrogenation.  M-GN is a semicondutor with indirect band gap of 0.64 eV, whereas fully-hydrogenated M-GN (named Me-graphane) is a wide band gap semiconductor with $E_{\text{g}}=2.81$ eV in GGA-PBE approach. Analyzing intermediate hydrogen concentrations for partial functionalization of M-GN we found metallic ground-states a semiconducting state for 15\%-hydrogenated M-GN, with narrow band gap of 0.11 eV. In Me-graphane, the effective masses of charge carriers is at least four times higher than those of pristine M-GN, although for 15\%-hydrogenated M-GN the effective electron mass almost halved and the effective hole mass is not significantly altered. The hydrogen atoms bind strongly to M-GN indicating chemisorption, with binding energies higher than 3 eV. Me-graphane presents higher bond lengths compared to M-GN ones, with boat-like conformer being its most favorable conformation. The reactive molecular dynamics simulation shows that the hydrogenation of M-GN is a temperature-dependent reaction, with formation of hydrogen islands starting with adsorptions in the most stable carbon site predicted in our DFT calculations. In conclusion, we believe our results will motivate the interest on the synthesis of M-GN and Me-graphane, what could boost the range of potential applications of carbon-allotropes.

\begin{acknowledgments}
This work was supported by the Brazilian agencies
FAPESP, CAPES, and CNPq (Process No. 310045/2019-3). Computational resources were provided by the high performance computing center at UFABC. The authors thank Mr. Matheus Medina for his technical support in reactive molecular dynamics simulations.
\end{acknowledgments}

%\appendix

%\section{Appendixes}

\bibliography{apssamp}% Produces the bibliography via BibTeX.

\end{document}